\documentclass[twoside,superscriptaddress,twocolumn,floatfix,a4paper,aps]{revtex4-1}

\usepackage{color}
\usepackage{graphicx}
\usepackage[utf8x]{inputenc}
\usepackage[T1]{fontenc}
\usepackage{siunitx}
\usepackage{amstext}
\usepackage{textgreek}
\usepackage{hyperref}
\usepackage{epstopdf}
\usepackage{amsfonts}
\usepackage{amsmath}
\usepackage{multirow}
\usepackage{dsfont}
\usepackage{physics}
\usepackage{filecontents}

\usepackage{marvosym}
\usepackage[rgb]{xcolor}

\makeatletter

\newcommand{\Rnum}[1]{\expandafter\@slowromancap\romannumeral #1@}

\begin{document}

\title{Experimental hybrid quantum-classical reinforcement learning by boson sampling:
how to train a quantum cloner}

\author{Jan Jašek} \email{jaseja00@upol.cz}
\affiliation{RCPTM, Joint Laboratory of Optics of Palacký University and Institute of Physics of Czech Academy of Sciences, 17. listopadu 12, 771 46 Olomouc, Czech Republic}

\author{Kateřina Jiráková} \email{katerina.jirakova@upol.cz}
\affiliation{RCPTM, Joint Laboratory of Optics of Palacký University and Institute of Physics of Czech Academy of Sciences, 17. listopadu 12, 771 46 Olomouc, Czech Republic}

\author{Karol Bartkiewicz} \email{karol.bartkiewicz@upol.cz}
\affiliation{RCPTM, Joint Laboratory of Optics of Palacký University and Institute of Physics of Czech Academy of Sciences, 17. listopadu 12, 771 46 Olomouc, Czech Republic}
\affiliation{Faculty of Physics, Adam Mickiewicz University,
PL-61-614 Pozna\'n, Poland}

\author{Antonín Černoch} \email{acernoch@fzu.cz}
\affiliation{Institute of Physics of the Czech Academy of Sciences, Joint Laboratory of Optics of PU and IP AS CR, 17. listopadu 50A, 772 07 Olomouc, Czech Republic}
  
\author{Tomáš Fürst}
\email{tomas.furst@upol.cz}
\affiliation{Department of Mathematical Analysis and Application of Mathematics, Faculty of Science, Palacký University, 17. listopadu 12, 771 46 Olomouc}

\author{Karel Lemr}
\email{k.lemr@upol.cz}
\affiliation{RCPTM, Joint Laboratory of Optics of Palacký University and Institute of Physics of Czech Academy of Sciences, 17. listopadu 12, 771 46 Olomouc, Czech Republic}   

\begin{abstract}
We report on experimental implementation of a machine-learned quantum gate driven by a classical control. The gate learns optimal phase-covariant cloning in a reinforcement learning scenario having fidelity of the clones as reward. In our experiment, the gate learns to achieve nearly optimal cloning fidelity allowed for this particular class of states. This makes it a proof of present-day feasibility and practical applicability of the hybrid machine learning approach combining quantum information processing with classical control. Moreover, our experiment can be directly generalized to larger interferometers where the computational cost of classical computer is much lower than the cost of boson sampling.
\end{abstract}

\date{\today}

\maketitle
\section*{Introduction}
\label{Introduction}

Machine learning methods are extensively used in an increasing number of fields, e.g., automotive industry, medical science, internet security, air-traffic control etc. This field conveys many algorithms and structures ranging from simple linear regression to almost arbitrarily complex (limited by computational resources and amount of training data) neural networks which are able to find solutions to highly nonlinear/complex problems. Recently, a considerable attention was drawn to the overlap between quantum physics and machine learning \cite{Biamonte2017}. Depending on the type of input data and data processing algorithms, we can distinguish four types of so called quantum machine learning (QML), i.e., CC (classical data and classical data processing -- classical limit of quantum machine learning), QC (quantum data and classical data processing), CQ (classical data and quantum data processing), and QQ (quantum data and quantum data processing).    

In addition to its fundamentally interesting aspects, QML can offer reduced computational complexity with respect to its classical counterpart in solving some classes of problems~\cite{Biamonte2017}. Depending on the problem at hand the speedup can be associated with various features of quantum physics. A number of proposals and experiments focused on QML have been reported, such works include for instance quantum support vector machines~\cite{Cai2015PRL}, Boltzmann machines~\cite{Gao2018PRL}, quantum autoencoders~\cite{PhysRevLett.122.060501}, kernel methods~\cite{Schuld2019PRL}, and quantum reinforcement learning~\cite{Hentschel2010,PhysRevLett.110.220501}. In reinforcement learning a learning agent receives feedback in order to learn an optimal strategy for handling a nontrivial task. Next, the performance of the agent is tested on cases that were not included in the training. If the agent performs well in these cases the validation is completed. 

\begin{figure}
\includegraphics[scale=0.33]{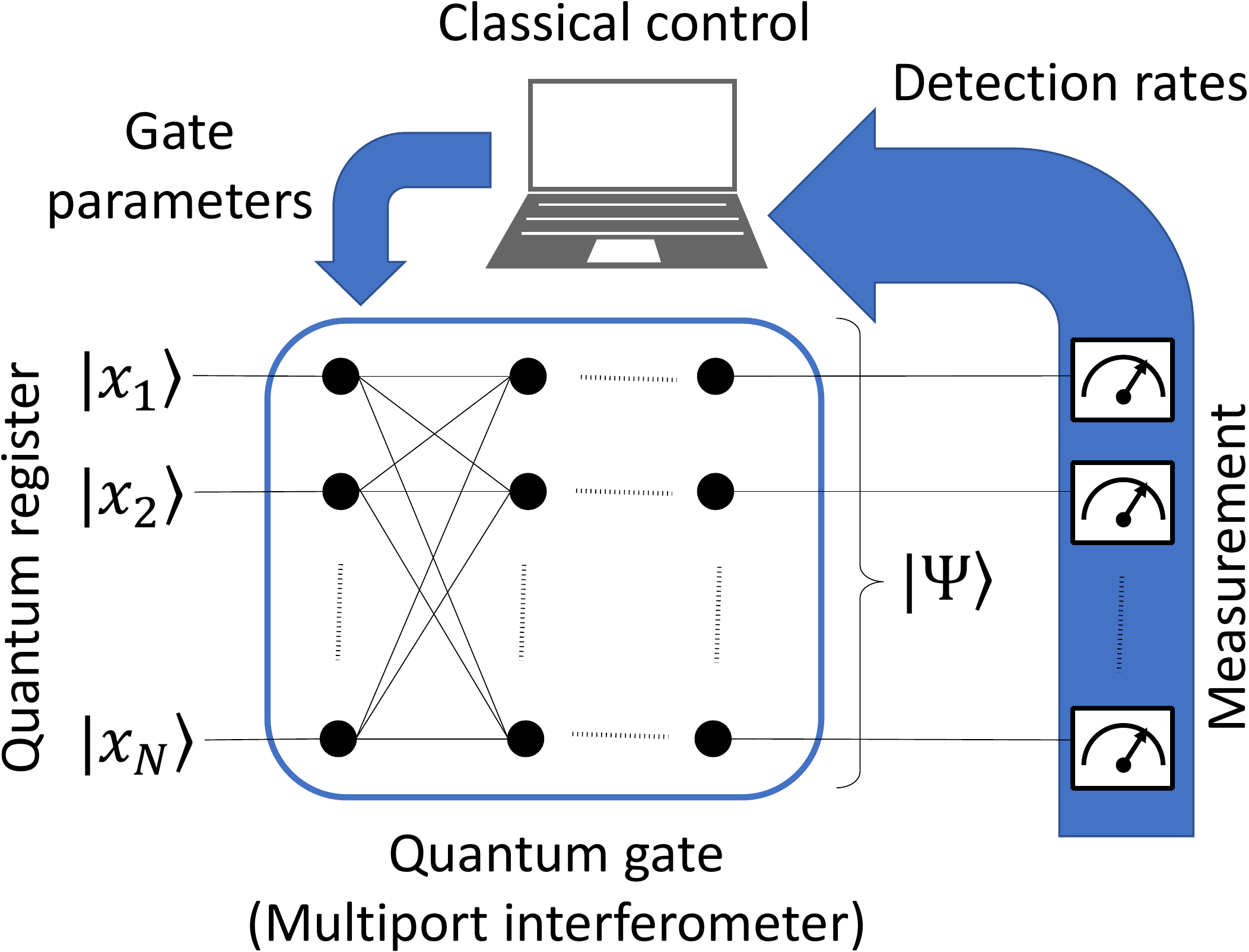}
\caption{\label{fig:concept} Conceptual scheme of hybrid reinforcement learning of a quantum gate driven by a classical control. The transformation of the quantum register performed by the gate is evaluated by measurement providing a reward to the classical control that iteratively modifies the gate's parameters.}
\end{figure}
In this paper, we demonstrate experimentally that reinforcement learning can be used to train an optical quantum gate (see conceptual scheme in Fig. \ref{fig:concept}). This problem is related to the boson sampling problem~\cite{BosonSampling,Broome2012,Spring2012,Tillmann2013,Crespi2013} where one knows the form of the scattering matrix of a system and learns its permanent. However, here we optimize the probabilities of obtaining certain outputs of the gate by finding the optimal parameters of the scattering matrix. Calculating the probabilities (permameters of scattering matrix) is in general a computationally hard task while measuring them is much faster. This feature of quantum optics allows us to expect that complex integrated interferometers could be applied as special-purpose quantum computers (for a 16 photon quantum interferometer see~\cite{Wang2018}). This sets our problem in the class of CQ quantum machine learning tasks. There are other QML approaches that could be used to optimize a quantum circuit. One approach uses classical machine learning to optimize the design of a quantum experiment in order to produce certain defined states~\cite{ODriscoll2018}. Another QML approach consists of optimizing quantum circuits to improve the solution to some problems solved on a quantum computer~\cite{Guerreschi2017}. The latter method is motivated by recent developments in quantum computing which suggest that using a quantum computer even to minor tasks within an experiment can save computational  resources~\cite{McClean2016,Peruzzo2014}. 

We applied online reinforcement learning methods to train an optimal quantum cloner. Quantum cloning is essential both for certain experimental tasks and for fundamental quantum physics. It is indispensable for safety tests of quantum cryptography systems or of other quantum communications protocols. Perfect quantum cloning of an unknown state is prohibited by the no-cloning theorem~\cite{Wootters1982}. However, it is possible to prepare imperfect clones that resemble the original state to a certain degree. Usually, the approach towards quantum cloning involves explicit optimization of the interaction between the system in the cloned state and another systems to maximize the fidelity of the output clones. Then, one uses such results explicitly to set the parameters of the experimental setup. In contrast to that, we present a quantum gate (learning agent) that is capable to self-learn such interaction (policy) based on provided feedback (implicit setting of the parameters). For the purposes of this proof-of-principle experiment, we limit ourselves to qubits in the form of
\begin{equation}
|\psi_{s}\rangle = 1/\sqrt{2} \left( |H\rangle + e^{i\eta} |V\rangle \right) \, , \label{eq:covstate}
\end{equation}
where $|0\rangle$ and $|1\rangle$ denote logical qubit states. These qubits can be found on the equator of the Bloch sphere, hence, we call them equatorial qubits. Cloning of these states is known to be the optimal means of attack on individual qubits of the famous quantum cryptography protocols BB84~\cite{Bennett1984} and RO4~\cite{Renes2004,Schiavon2016} or quantum money protocol~\cite{Wiesner83} (see also Refs.~\cite{Bartkiewicz13PRL,Bartkiewicz2017} for experimental implementation).

\section*{Results}
\label{Results}

We demonstrate reinforcement-learned quantum cloner for a class of phase-covariant quantum states. For the training procedure, the figure of merit is the individual fidelity of the output copies. The fidelity of the $j$-th clone is defined as overlap between the state of the input qubit $|\psi\rangle_{\text{in}}$ and the state of the clone $\hat{\varrho}_{j}$:
\begin{equation}
F_{j} ={}_{\text{in}}\langle \psi|\hat{\varrho}_{j}|\psi\rangle_{\text{in}}\, . \label{eq:fidelita}
\end{equation}
In case of the state in Eq.~(\ref{eq:covstate}), the maximum achievable fidelity of symmetrical $1 \rightarrow 2$ cloning accounts for $F_{1} = F_{2} = \frac{1}{2} \left(1+\frac{1}{\sqrt{2}}\right) \approx 0.8535
$~\cite{Fiurasek2003,Bartkiewicz2009}.

We have constructed a two qubit gate on the platform of linear optics. Qubits were encoded into polarization states of the individual photons ($|0\rangle \Leftrightarrow |H\rangle$ and $|1\rangle \Leftrightarrow |V\rangle$). The gate operates formally as a polarization dependent beam splitter with tunable splitting ratios for horizontal ($H$) and vertical ($V$) polarization. This tunability provides two parameters for self-learning labeled $\phi$ and $\theta$ throughout the text. The third learnable parameter $\omega$ is embedded in the state of the ancillary photon
\begin{equation}
\ket{\psi_\text{a}} = \cos{2\omega} \ket{H} + \sin{2\omega} \ket{V}\, . \label{eq:ancilla}
\end{equation}
%
\begin{figure*}
\includegraphics[scale=1]{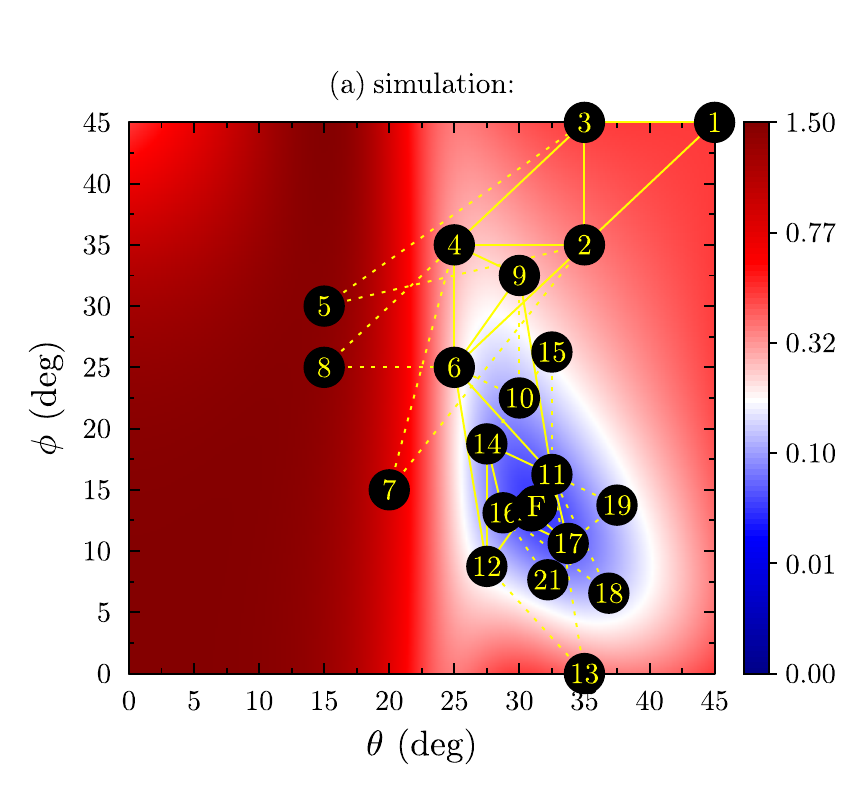}\hspace{2em}
\includegraphics[scale=1]{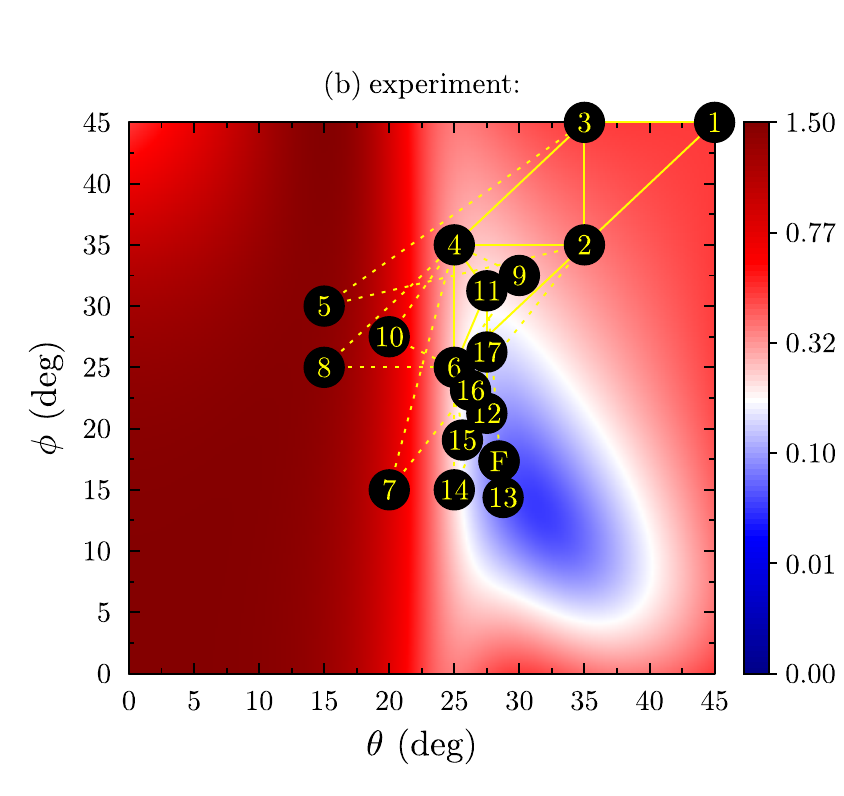}
\caption{\label{fig:potencial} Plot of the cost function $C$ for the angles $\phi$ and $\theta$ (corresponding to $\text{HWP}_{4}$ and $\text{HWP}_{3}$ in Figure~\ref{fig:Exsetup}, respectively). The solid yellow lines denote the final triangles reached by the Nelder--Mead simplex minimization \cite{Nelder1965} algorithm in each of its iteration. The dashed lines mark intermediate steps. The circled numbers stand for gate runs and point F depicts the final state of the gate at the end of training. A simulation was performed prior to the experiment to verify our implementation of the learning algorithm. Differences between the experimentally learned and simulated parameters can be explained by imperfections in the experimental setup.}
\end{figure*}
%

\begin{figure*}
\includegraphics[scale=1]{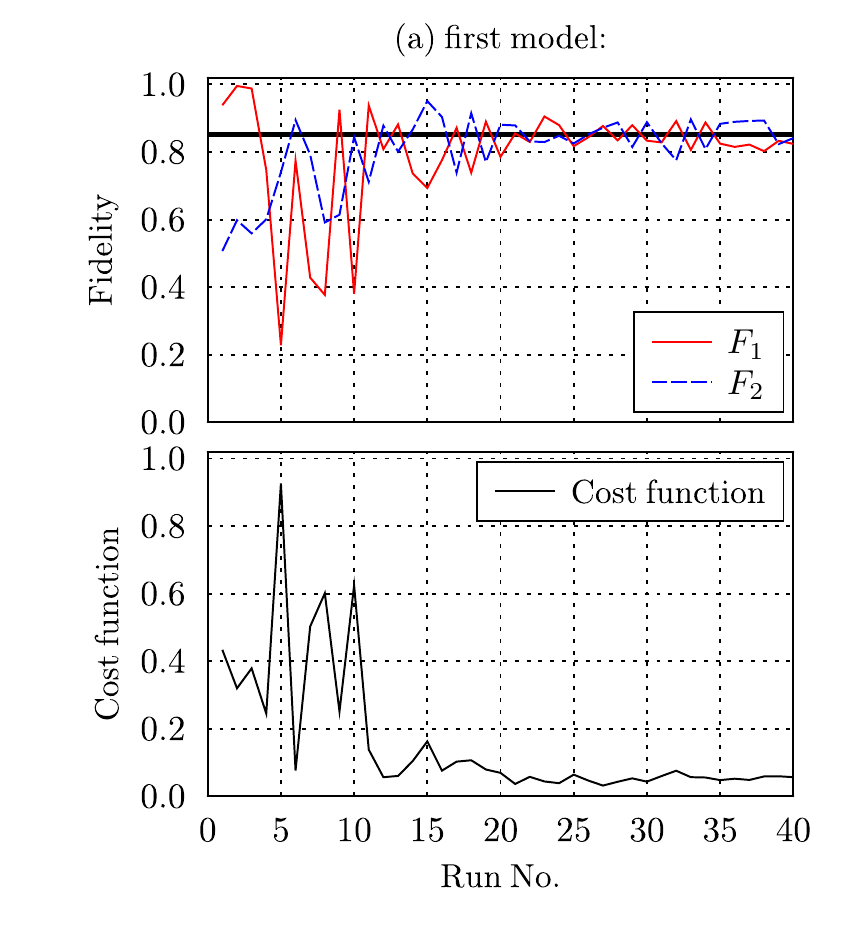}
\includegraphics[scale=1]{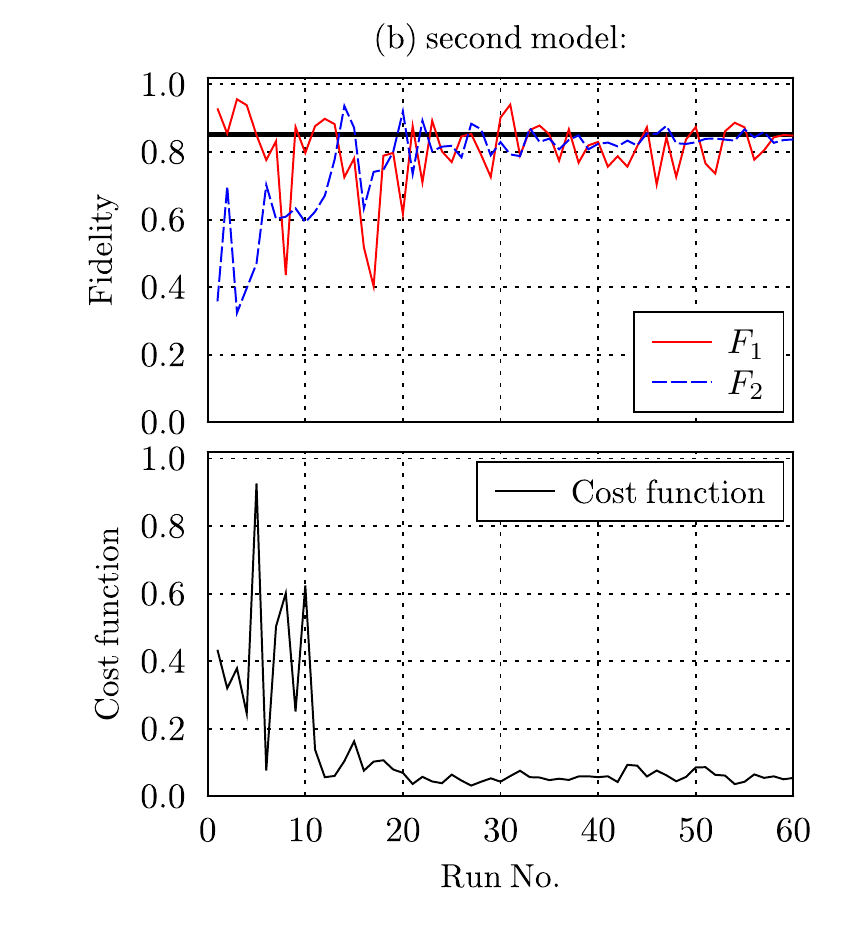}
\caption{\label{fig:fid} Plots showing the evolution of fidelity of both the clones (top) as well as the cost function (bottom) throughout the training in case of (a) the first model with two free parameters $\phi$ and $\theta$ and (b) the second model with three free parameters $\phi$, $\theta$ and $\omega$. Fidelity of the first clone $F_{1}$ is visualized by a solid red line and the fidelity of the second clone $F_{2}$ is shown in blue (dashed line). The thick solid black line stands for the theoretical limit of $\approx 0.8535$.}       
\end{figure*}

We have experimentally implemented two machine learning models using two and three parameters, respectively. In the first model, we fixed the ancilla state to its theoretically known optimum $\ket{\psi_\text{a}} = \ket{H}$. The remaining two parameters of the gate $\phi$ and $\theta$ were machine learned. To minimize the cost function (i.e. optimize the performance of the cloner) we applied Nelder-Mead simplex algorithm which iteratively searches for a minimum of a cost function. We chose the cost function to be in the form of
\begin{equation}
C = (1-F_{1})^2+(1-F_{2})^2+(F_{1}-F_{2})^2\, ,
\end{equation}
where $F_{1}$ and $F_{2}$ stand for the fidelity of the first and second clone, respectively. This choice reflects the natural requirements to obtain maximum fidelities of both the clones as well as to force the cloner into a symmetrical cloning regime. Training of the gate consists of providing it with training instances of equatorial qubit states (randomly generated in each cost function evaluation, i. e. an online machine learning scenario) and with the respective fidelity of the clones. In each training run, the underlying Nelder-Mead algorithm sets the gate parameters to vertices of simplexes in the parameter space and then decides on a future action. In the case of a two-parameter optimization, these simplexes correspond to triangles as depicted in  Figure~\ref{fig:potencial}. In this Figure, we plot the exact path taken by the Nelder-Mead simplex algorithm to minimize the cost function $C$ for the case of a real experiment and its simulation. The selected initial simplex was intentionally chosen well away from the optimal position -- its first vertex resembles the trivial cloning strategy ~\cite{Bartkiewicz2014,Bartkiewicz2017}. In Figure~\ref{fig:fid}a, we illustrate the evolution of both the fidelities $F_{1}$ and $F_{2}$ during the training. After 40 runs (i.e. 40 instances from the training set), this model was deemed trained because the size of simplexes dropped to the experimental uncertainty level (i.e. $\sim 0,1°$ on rotation angles of wave plates). However, in general, setting the simplex to converge within a given precision is a nontrivial problem~\cite{nash1990compact}.

In the second model, we let the gate learn the optimal setting of the ancilla $\omega$ along with the gate parameters $\phi$ and $\theta$. The training procedure ran similarly to the first model. The initial value of the $\omega$ parameter was set naively to $\omega = \tfrac{\pi}{8}$ so it lied on the equator of the Poincaré-Bloch sphere. We present evolution of the intermediate fidelities of this three-parameter model in Figure~\ref{fig:fid}b. Using a similar stopping criterion as in the first model, the training of the second model was terminated after 60 runs.

We have tested the performance of both our models on independent random test sets each populated by 40 instances of equatorial states. We summarize the results of the two models in Table~\ref{tab:sum}, where we provide the final learned parameters together with the mean values of the fidelities on the test sets $\langle F_{1} \rangle$ and $\langle F_{2} \rangle$. The observed fidelities on test sets are bordering on the theoretical limit (at most 0.013 below it) which renders our gate highly precise in context of previously implemented cloners~\cite{Soubusta2007,Soubusta2008,Bruss00,Fiurasek2003,Xu2008,Buscemi2005}.

\begin{table}
\begin{ruledtabular}
\caption{Summary of the final values for both models. $\langle F_{1} \rangle$ and $\langle F_{2} \rangle$ denote the mean fidelities observed on the test sets.}
\label{tab:sum}
\begin{tabular}{lll}
Final values     & Model 1 & Model 2 \\
 \hline
\sl{training:} & & \\
Angle $\phi$     &     28.59         &     27.44         \\
Angle $\theta$   &     17.11         &      21.68        \\
Angle $\omega$   &    45° (Fixed)          &      41.49°   \\
No. of training   &     40                      &      60         \\
instances (runs)             &                             &               \\
\hline
\sl{testing:} & &\\
$\langle F_{1} \rangle$ &      $0.840 \pm 0.033$        &       $0.849 \pm 0.040$        \\
$\langle F_{2} \rangle$ &      $0.843 \pm 0.046$         &      $0.853 \pm 0.022$

\end{tabular}
\end{ruledtabular}
\end{table} 

\section*{Experimental realization}
\label{ch:experiment}

\begin{figure}
\centering
\hfill\includegraphics[scale=0.7]{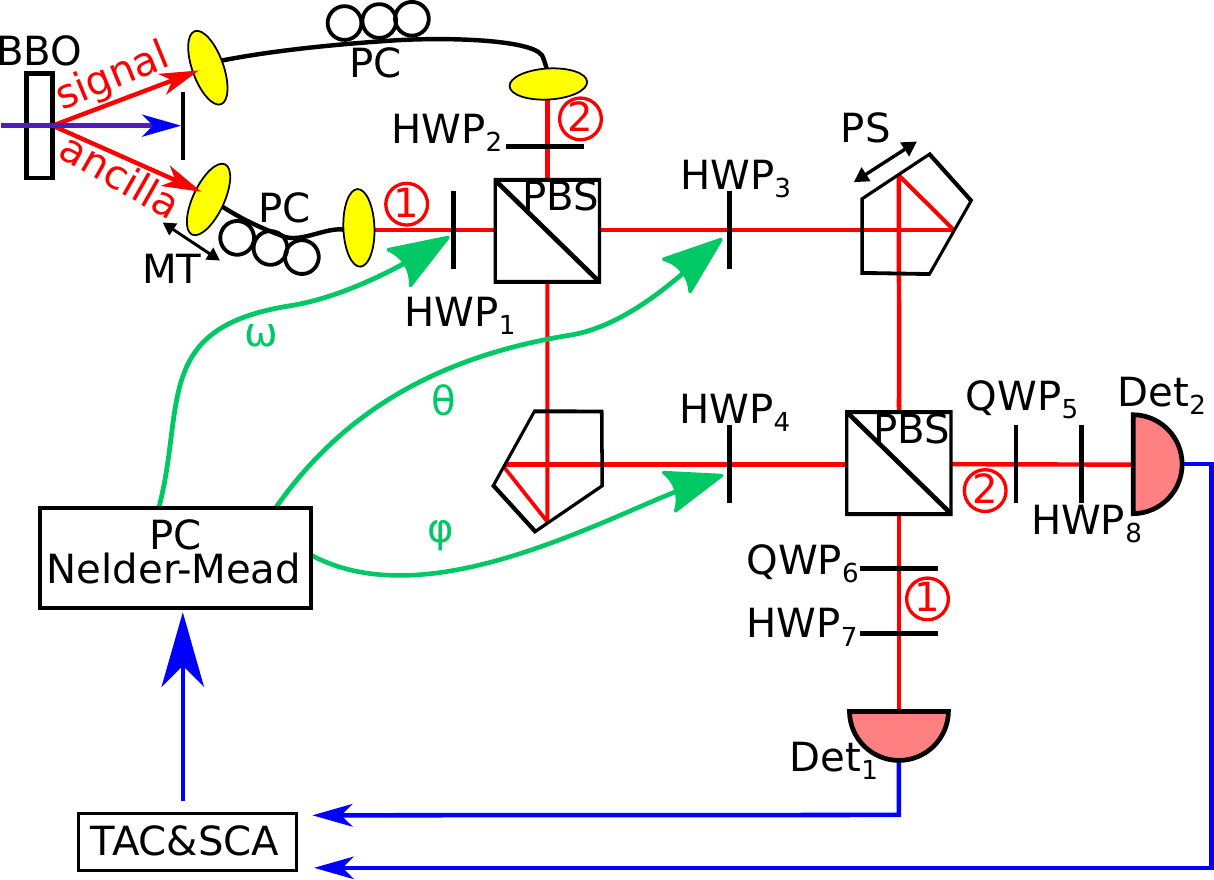}\hspace*{\fill}
\caption{Experimental setup. Legend: PBS -- polarization beam splitter, PC -- polarization controller, BBO -- barium beta borate, Det -- detector, HWP -- half-wave plate, QWP -- quarter wave-plate, PS -- piezoelectric stage, TAC$\&$SCA -- time-to-amplitude converter \& single channel analyzer.}
\label{fig:Exsetup}
\end{figure}

We constructed a device composed of a linear optical quantum gate and a computer performing classical information processing. While the gate itself is capable of a broad range of two qubit transformations, this paper focuses on its ability to act as a phase-covariant quantum cloner. The experimental setup is depicted in Fig. \ref{fig:Exsetup}. Pairs of photons are generated in Type I spontaneous parametric down-conversion occurring in a nonlinear BBO crystal. This crystal is pumped by Coherent Paladin Nd-YAG laser with integrated third harmonic generation of wavelength at $\lambda = 355$ nm. The generated pairs of photons are both horizontally polarized and highly correlated in time. 

These photons are then spectrally filtered by 10 nm wide interference filters and spatially filtered by two single mode optical fibers each guiding one photon from the pair. In our experimental setup, qubits are encoded into polarization states of the individual photons. The photon in the upper path (spatial mode 2) represents the signal qubit, quantum state of which we want to clone, and the photon in the lower path (spatial mode 1) serves as the ancilla.  

Using polarization controllers (PC) we can ensure that both photons are horizontally polarized at the output of the fibers. From now on, polarization states of the photons are set using a combination of half-wave plates (HWPs) and quarter-wave plates (QWPs). There are eight wave plates in total, two stationary QWPs fixed at angle 45\textdegree{} and six motorized HWPs which make it possible to control the whole quantum gate using a computer. The first two half-wave plates HWP$_1$ and HWP$_2$ are used to set input polarisation states of the photons.

The core part of the presented quantum gate is a Mach-Zehnder-type interferometer which consists of two polarizing beam splitters (PBS) and two reflective pentaprisms, one of which is attached to the piezoelectric stage (PS). With the addition of two HWPs (HWP$_3$ and HWP$_4$) placed in its arms, this whole interferometer implements a polarization dependent beam splitter with variable splitting ratio. Mathematically, the scattering matrix of the gate reads

\begin{align}
\begin{split}
\begin{pmatrix}
\hat{a}_{\text{H},1} \\ \hat{a}_{\text{V},1} \\ \hat{a}_{\text{H},2} \\ \hat{a}_{\text{V},2} 
\end{pmatrix} \rightarrow
\begin{pmatrix}
\cos 2\phi & 0 & \sin 2\phi & 0 \\
0 & \cos 2\theta & 0 & \sin 2\theta \\
- \sin 2\phi & 0 & \cos 2\phi & 0 \\
0 & - \sin 2\theta & 0 & \cos 2\theta \\
\end{pmatrix}
\begin{pmatrix}
\hat{a}_{\text{H},1} \\ \hat{a}_{\text{V},1} \\ \hat{a}_{\text{H},2} \\ \hat{a}_{\text{V},2} 
\end{pmatrix}
\label{eq:beamsplitmodel1}
\end{split}
\end{align} where $\hat{a}_{x,i}$ represents the annihilation operators of the individual input, polarization ($x \in \left\lbrace \text{H,V} \right\rbrace $) as well as spatial ($i \in \left\lbrace \text{1,2} \right\rbrace $)  modes. The angles $\theta$ and $\phi$ correspond to the rotations of HWP$_3$ and HWP$_4$ with respect to the horizontal axis. The transformation (\ref{eq:beamsplitmodel1}) is formally equivalent to the transformation by a polarization dependent beam splitter, the intensity splitting ratios of which for horizontal and vertical polarizations are $ \cot^2{2\phi} $ and $ \cot^2{2\theta}$, respectively. 

The two spatial modes at the output of the interferometer are subjected to polarization projection (QWP$_5$, QWP$_6$ and HWP$_7$, HWP$_8$) and then led via single-mode optical fibers to a pair of avalanche photodiodes by Perkin-Elmer running in Geiger mode. We use detection electronics to register both single photons at each of the detectors and coincident detections as successful operation of the gate is indicated by the presence of single photon in each output of the interferometer. The electronic signal is then sent to a classical computer.

For specific parameters of the presented linear-optical elements, this quantum gate functions as a $1 \rightarrow 2$ symmetric phase-covariant cloner, optimal analytical cloning transformation of which is well known~\cite{Fiurasek2003}. On a linear-optical platform, this optimal cloning transformation can be achieved by a polarization dependent 
beam splitter with intensity transmitivities for horizontal and vertical polarization at $t_\text{H} \approx 0.19 $ and $t_\text{V} \approx 0.81 $, while setting the ancilla to be horizontally polarized. Note that our quantum gate is capable of implementing this transformation when set approximately to $\phi = 31.3, \theta = 13.7$ and the maximum theoretical fidelities of the clones are  $F_1 = F_2 = \frac{1}{2} \left( 1 + \frac{1}{\sqrt{2}} \right) \approx 0.8535$.

To showcase the capability of our gate to learn to clone phase-covariant states optimally, we deliberately ignore this analytical solution and employ self-optimization procedure seeking to maximize the cloning fidelities. The optimization process consists of a number of measurements (runs), each performed for a set of variable optimization parameters $\phi, \theta, \omega$. That is, variable splitting ratio for horizontal ($\phi$) and vertical ($\theta$) polarization as well as the state of the ancilla ($\omega$) (Eq.~\ref{eq:ancilla}) controlled by the rotation of HWP$_1$. In each run, output clones fidelities are evaluated and supplied to the classical Nelder-Mead algorithm for a decision about the parameters of the future runs. 

In between any two runs, the setup is stabilized. We first minimize temporal delay between the two individual photons. In this case, all HWPs are set to 0\textdegree{} with the exception of HWP$_4$ being at 22.5\textdegree{}. In this regime, we minimize the number of two-photon coincident detections (Hong-Ou-Mandel dip) by changing the temporal delay between the photons using a motorized translation stage MT. In the next step, the phase is stabilized in the interferometer. Moreover, we make use of the fact that the phase shift in the interferometer additively contributes to the phase $\eta$ of the signal state (Eq.~\ref{eq:covstate}). This allows us to use interferometer phase stabilization for setting of any signal state of the equatorial class. We achieve this task by setting HWP$_2$ to 22.5\textdegree{} and HWP$_8$ to the value corresponding to orthogonal state with respect to the required input signal state. All other HWPs are set to 0\textdegree{} and a minimum in single-photon detections on Det 2 is found by tuning the voltage applied to PS. Note that the entire stabilization procedure is completely independent of the learning process itself.

While all six of the HWPs are controlled by a PC, only three (HWP$_1$, HWP$_3$ and HWP$_4$) are specifically controlled by the optimization algorithm. In contrast to that, HWP$_2$ is used to set the quantum state of the cloned qubit and HWP$_7$ and HWP$_8$ are used to choose polarization projections, therefore their configuration shall not be accessible to the optimization algorithm.

In this reinforced-learning scenario, the cloner is trained on a sequence of random equatorial signal states (Eq.~\ref{eq:covstate}) different for each run. The phase $\eta$ is randomly picked from interval $\left( 0; 2 \pi \right)$. The optimization algorithm then rotates HWP$_1$, HWP$_3$ and HWP$_4$ to chosen angles $\phi , \theta , \omega$. Finally, the cloner is fed back the measured fidelities $F_1, F_2$ of the clones.

The fidelities are obtained by measuring coincidence detections in four different projection settings that correspond to the angles set on HWP$_7$ and HWP$_8$. We label these coincident detections $cc_{ij}$, where $i,j \in \left\lbrace \parallel; \perp \right\rbrace$. The $\parallel$ and the $\perp$ sign denote projection on the signal state $\ket{\psi_\text{s}}$ and its orthogonal counterpart $\ket{\psi_\text{s}^\perp}$. We calculate the fidelities as
\begin{equation}
F_1 = \frac{cc_{\parallel\parallel} + cc_{\parallel\perp}}{\Sigma} , F_2 = \frac{cc_{\parallel\parallel} + cc_{\perp\parallel}}{\Sigma},
\end{equation} where $\Sigma$ denotes $cc_{\parallel\parallel} + cc_{\parallel\perp} + cc_{\perp\parallel} + cc_{\perp\perp}$.

The core part of the optimization process is the Nelder-Mead simplex algorithm which minimizes a chosen cost function $C$ in a multidimensional space corresponding to the number of the function parameters $N$~\cite{Nelder1965}. The algorithm takes $\left( N+1 \right)$ points in the parameter space to create a $\left( N+1 \right)$ dimensional initial simplex (each point corresponds to one of the simplex vertices). For example, with 2 parameters being optimized, the algorithm creates a triangle, for 3 parameters it creates a tetrahedron and so on. The value of the cost function at each point of the simplex is then evaluated and the algorithm transforms the simplex in such way to find a point of local minimum. In our case, we define the cost function $C$ so that the first two elements maximize the obtained fidelities, while the last element achieves symmetry of the cloning.

\section*{Conclusions}
\label{ch:concl}
In our proof of principle experiment, we implemented a CQ reinforcement quantum machine learning algorithm driven by a hybrid of classical Nelder--Mead method and quantum computing based permanent measurement. This approach was used to train a practical quantum gate (i.e., a quantum cloner). The task of the training was to optimize parameters of the gate (interferometer) $\phi$, $\theta$ and $\omega$ (setting of the ancilla in the second experiment) to perform phase-covariant cloning. The quality of both the clones measured by their fidelities $F_{1}$ and $F_{2}$ which were evaluated within both experiments (Figure~\ref{fig:fid}) successfully reached the theoretical limit for phase-covariant cloning $0.854$. Remarkably, the cloner managed to achieve almost optimal cloning by learning setup parameters, slightly different from analytical values, that counter all experimental imperfections including imperferctions in the cloner itself and in the input state preparation.

To see the connection between boson sampling and our results, let us focus on computing the permanent $\mathrm{Perm}$ of scattering matrix describing the gate operation. The unitary scattering matrix $U$ performs linear transformation on the annihilation operators $a_i$ of the input modes ($i$ can be an index labeling both the polarization and the spatial degrees of freedom). Then the input-output relation of an quantum-optical interferometer is given as $b_j = \sum U^\dagger_{i,j} a_i.$ If all the input modes of an interferometer are injected with single photons and single photons are detected at specific outputs (no bunching) the probability of obtaining the desired detection coincidence is $p = |\mathrm{Perm}U|^2.$ However, this expression becomes more complex if some modes are occupied by more than one photon. Then factorials of mode-specific photon numbers appear as denominator and the respective rows/columns of $U$ must be repeated a corresponding number of times \cite{BosonSampling}. If some output modes are not to be populated, the respective row of $U$ matrix is deleted. Calculating the permanents of the scattering matrix associated with our cloner by hand is already challenging (we have polarization and spatial degrees of freedom for two photons) and in general it falls into the $\#P$-hard complexity class. Our experiment can be directly generalized to larger interferometers, where the computational cost of classical computer is much lower than the cost of boson sampling. This makes our research a relevant application of so-called quantum circuit learning described in Ref.~\cite{PhysRevA.98.032309}.

Our results also opens possibilities of further research or applications in the field of quantum key distribution. Suppose a typical attack on the key distribution scheme: Bob and Alice share quantum states and the attacker Eve is eavesdropping on them. Bob and Alice exchange quantum states and, via a classical line, they can decide to stop exchanging qubits (because of noise). Let us assume that Eve is eavesdropping on both quantum and classical communication. Eve can in principle use reinforced-learning to train a cloner to perform the attack by feeding it with information on the behavior of Bob and Alice, e.g., their decision on  continuing or aborting the exchange of a quantum key and/or their decision on parameters of privacy amplification. For such application the proposed gate would have to be modified since Eve does not know the specific class of states used by Bob and Alice, but that is out of the scope of this paper.

\vspace{1em}
\begin{acknowledgments}
K.J., K.B., A.Č. and K.L. acknowledge financial support by the Czech Science Foundation under the project No. 19-19002S. The authors also acknowledge project CZ.02.1.01/0.0/0.0/16\_019/0000754 of the Ministry of Education, Youth and Sports of the Czech Republic. J.J. and K.J. also acknowledge the Palacky University internal grant No. IGA-PrF-2019-008 and T.F. gratefully acknowledges the support by the Operational Programme Research, Development and Education, project no. CZ.02.1.01/0.0/0.0/17\_049/0008422 of the Ministry of Education, Youth and Sports of the Czech Republic.
\end{acknowledgments}
\bibliography{citace}

\end{document}